\documentclass[12pt,a4paper,twoside,english]{article}
\usepackage[T1]{fontenc}
\usepackage[latin9]{inputenc}
\usepackage{float}
\usepackage{graphicx}
\usepackage{setspace}
\makeatletter

\pdfpageheight\paperheight
\pdfpagewidth\paperwidth

\newcommand{\noun}[1]{\textsc{#1}}
\providecommand{\tabularnewline}{\\}

\newcommand{\lyxaddress}[1]{
\par {\raggedright #1
\vspace{1.4em}
\noindent\par}
}

\@ifundefined{date}{}{\date{}}
\makeatother

\usepackage{babel}
\begin{document}

\title{Two different modes of oscillation in a gene transcription regulatory
network with interlinked positive and negative feedback loops}

\author{Rajesh Karmakar%
\thanks{Email: rkarmakar2001@yahoo.com%
}}

\maketitle

\lyxaddress{\begin{center}
Department of Physics, AKPC Mahavidyalaya, Subhasnagar, Bengai, Hooghly,
PIN-712611, W.B., INDIA
\par\end{center}}
\begin{abstract}
We study the oscillatory behaviour of a gene regulatory network with
interlinked positive and negative feedback loop. Frequency and amplitude
are two important properties of oscillation. Studied network produces
two different modes of oscillation. In one mode (mode 1) frequency
remains constant over a wide range amplitude and in other mode (mode
2) the amplitude of oscillation remains constant over a wide range
of frequency. Our study reproduces both features of oscillations in
a single gene regulatory network and show that the negative plus positive
feedback loops in gene regulatory network offer additional advantage.
We identified the key parameters/variables responsible for different
modes of oscillation. The network is flexible in switching between
different modes by choosing appropriately the required parameters/variables.
\end{abstract}

\section{Introduction}

The fundamental unit of life is the cell. Organisms may consist of
just one cell or they may be multicellular. The multicellular organisms
organized into tissues which are groups of similar cells arranged
so as to perform a specific function. One may view cell life as a
collection of networks interacting through proteins, RNA, DNA and
small molecules involved in signaling and energy transfer. These networks
process environmental signals, induce cellular responses and execute
internal events such as gene expression, thus allowing cells and entire
organisms to perform their basic functions. These control and communication
networks can be relatively simple (in bacteria) or they may be incredibly
sophisticated (in higher organisms). In addition to their own needs
for survival and reproduction, cells in multicellular organisms need
additional levels of complexity in order to enable communication among
cells and overall regulations. In living organism, proteins are the
functional molecules. They are synthesized in a regulated processes
known as Gene Expression (GE). So, gene expression and regulation
are of fundamental importance in cell. Again, proteins from one gene
regulate the expression from other. In this way, gene regulatory networks
have grown inside the cell. There can be other type of networks like
metabolic networks, protein-protein interaction networks etc$^{1}$.
In general, the structure or architecture of the networks determines
the function of the networks$^{2}$. It is observed that positive
and negative feedback loops are very common motifs in biological networks.
They occur frequently in different gene regulatory and cell signaling
circuits. In general, it is known that positive feedback loop induces
a switch like behaviour and bistability and that negative feedback
loop produces oscillations, suppresses noise/fluctuation effects etc.
The loops are often coupled to perform various functions in the networks
acting as bistable switches, oscillators, excitable devices etc.$^{3-9}$. 

Rhythmic phenomena represent one of the most striking manifestations
of dynamic behaviour in biological systems$^{10-13}$. Cellular rhythms
are generated by complex interactions among genes, proteins and metabolites.
They are used to control signaling, motility, growth, division and
death. These rhythms appear in many regulatory mechanisms that control
the dynamics of living system. For example, neural and cardiac rhythms
are associated with the regulation of voltage dependent ion-channels,
metabolic oscillations originate from the regulation of enzyme activity
and intracellular calcium oscillations involve the control of transport
process while regulation of gene expression underlies circadian rhythms
at the cellular level. There are some essential requirements for biochemical
oscillations$^{14-15}$. In the course of time, open systems that
exchange matter and energy with their environment generally reach
a stable steady state. However, once the system operates sufficiently
far from equilibrium and when its kinetics acquires a sufficient nonlinearity,
the steady state may become unstable. Feedback processes and cooperativity
are two main sources of nonlinearity that favour the occurrence of
instabilities in biological systems. When the steady state becomes
unstable, the system moves away from it, often bursting into sustained
oscillations around the unstable steady state. Theoretical analysis
shows that a negative feedback networks with sufficient amount of
time delay and nonlinearity produces oscillations$^{4,11,13}$. The
time delay in the networks can be created by a long chain of intermediate
reactions or by an extra positive feedback loop. Different types of
interlinked positive and negative feedback loops are observed in cellular
systems with different number of nodes and links (Fig. 1). Such coupled
loops play a variety of roles, acting as bistable switches, oscillators
etc., although a single positive and negative feedback loop can also
perform these functions under certain conditions. It is demonstrated
that coupled or interlinked feedback loops are superior to single
feedback loops as oscillators$^{16}$. A superior oscillator has the
property of constant amplitude over a wide range of frequency. There
may be another type of oscillation in which frequency remains constant
though amplitude of oscillations may varied. Constant amplitude oscillations
are important in heart beat, cell cycle etc. For circadian oscillations
frequency should remain constant in different environmental conditions.
Here we study a gene regulatory network which show both type oscillations
depending on the variation of appropriate parameter.

\section{Interlinked gene transcription regulatory network: The model}

We consider a gene regulatory network consist of three genes X, Y
and Z which synthesizes three proteins $x$, $y$ and $z$ respectively.
Three genes form a closed loop structure and the product of each gene
represses the synthesis process from other in a cyclic way starting
from X to Y to Z. In addition to that there is a autocatalytic positive
feedback loop in X. The network architecture is identical to the module
considered by Tsai et al.$^{16}$. Only difference is that, the network
module considered by Tsai et al. is regulated at the degradation level
but in our network the regulation is achieved at the synthesis level.
The notwork is shown in Fig. 1. The dynamics of the network is driven
by the following coupled nonlinear differential equation.

\begin{equation}
\frac{dx}{dt}=-k_{2}\, x+\frac{k_{1}}{K_{1}^{n_{1}}+z^{n_{1}}}+\frac{k_{7}\: x^{n_{4}}}{K_{4}^{n_{4}}+x^{n_{4}}}
\end{equation}

\begin{equation}
\frac{dy}{dt}=-k_{4}\, y+\frac{k_{3}}{K_{2}^{n_{2}}+x^{n_{2}}}
\end{equation}

\begin{equation}
\frac{dz}{dt}=-k_{6}\, z+\frac{k_{5}}{K_{3}^{n_{3}}+y^{n_{3}}}
\end{equation}

The equations contain basically three different kinds of terms, viz.,
degradation, negative transcription or repression and autocatalysis.
The oscillatory behaviour (Figs. 2 and 3) of the interlinked gene
transcription regulatory network is studied by varying the different
rate constants. We solve the coupled nonlinear equations to observe
oscillation numerically by Runge-Kutta 4 technique. To verify the
stability of the network, we consider the random parameter values
in the range given in Table 1. We observe that 500 out of 8404 parameter
sets (5.94\%) yielded the oscillations in presence of positive feedback
loop. But in absence of autocatalytic positive feedback loop we observe
that 500 out of 7937 parameter sets (6.3\%) yielded the oscillations.
The last result (network without autocatalytic feedback loop) is completely
different from the result of Tsai et al.$^{16}$. To study the role
of positive feedback loop in the network we measure frequency and
amplitude of oscillations from the 500 oscillatory data sets with
different rate constants as variable. Then we take a particular set
of rate constant chosen from the 500 sets of parameter values for
which oscillations are observed. When we vary the repression strength
(k$_{1}$) on X, we observe that frequency remains constant over a
wide range of amplitude. As the positive feedback strength increases
the range of amplitude of oscillation over which the frequency remains
constant increases (Fig. 4). But if we vary the degradation rate constant
(k$_{2}$), we observe that amplitude remains constant over a wide
range of frequency. The autocatalytic loop in X increases that behaviour
further (Fig. 5). Same behaviour is observed when both k$_{1}$ and
k$_{2}$ varies simultaneously (Fig. 6). This observation shows that
the degradation rate has more impact on the oscillatory behaviour
of the network. Fig. 7 shows that amplitude remains constant over
a wide range of frequency when varied the autocatalytic positive feedback
strength k$_{7}$.

\begin{figure}[H]
\begin{centering}
\includegraphics[width=1.5in]{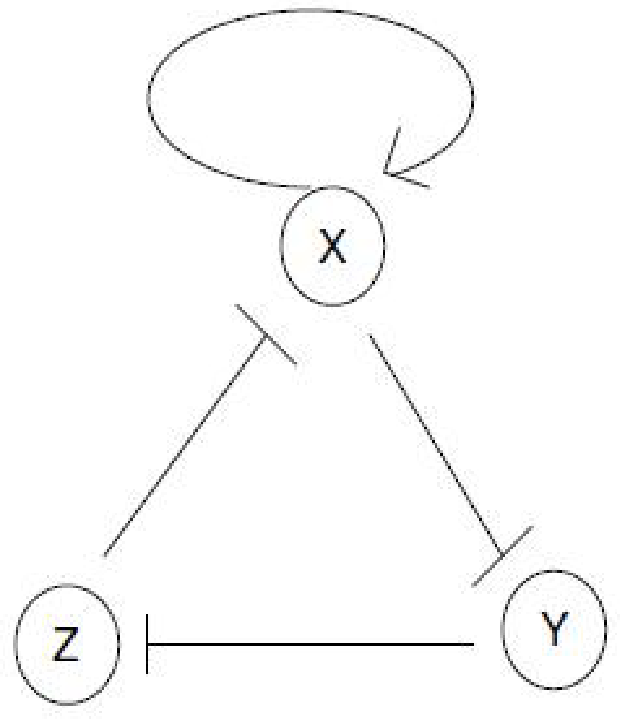} 
\par\end{centering}

\begin{centering}
Fig. 1. Gene transcription regulatory network with interlinked positive
and negative feedback loop.
\par\end{centering}

\end{figure}

\begin{figure}[H]
\begin{centering}
\includegraphics[width=2in]{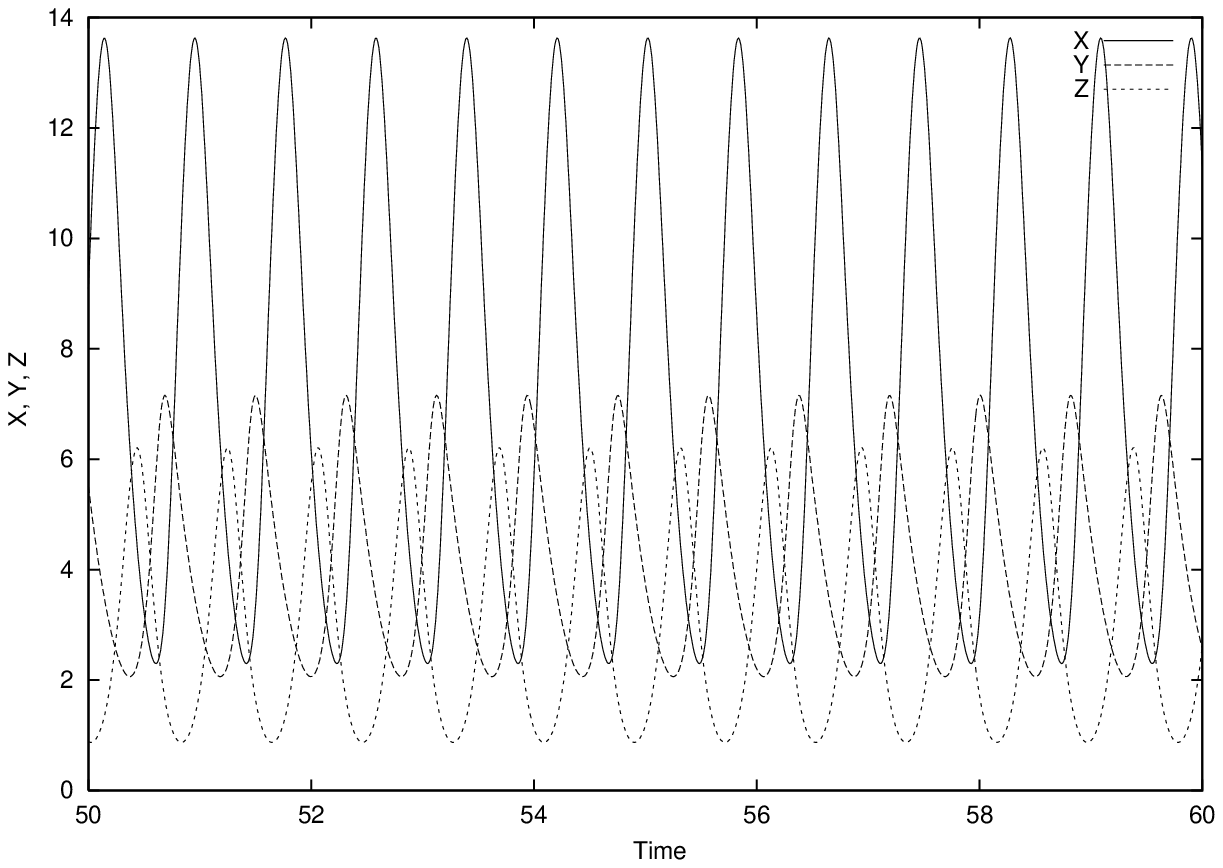}
\par\end{centering}

\begin{centering}
Fig. 2. Time dependent Oscillatory behaviour of X, Y and Z for the
parameter values k$_{1}$=266.152, k$_{2}$=5.730, k$_{3}$=331.660,
k$_{4}$=3.681, k$_{5}$=494.232, k$_{7}$=9.168, K$_{1}$=1.280,
K$_{2}$=0.982, K$_{3}$=0.959, K$_{4}$=18.882, n$_{1}$=2.658, n$_{2}$=2.048,
n$_{3}$=2.512, n$_{4}$=3.940. 
\par\end{centering}

\end{figure}

\begin{figure}[H]
\begin{centering}
\includegraphics[width=2in]{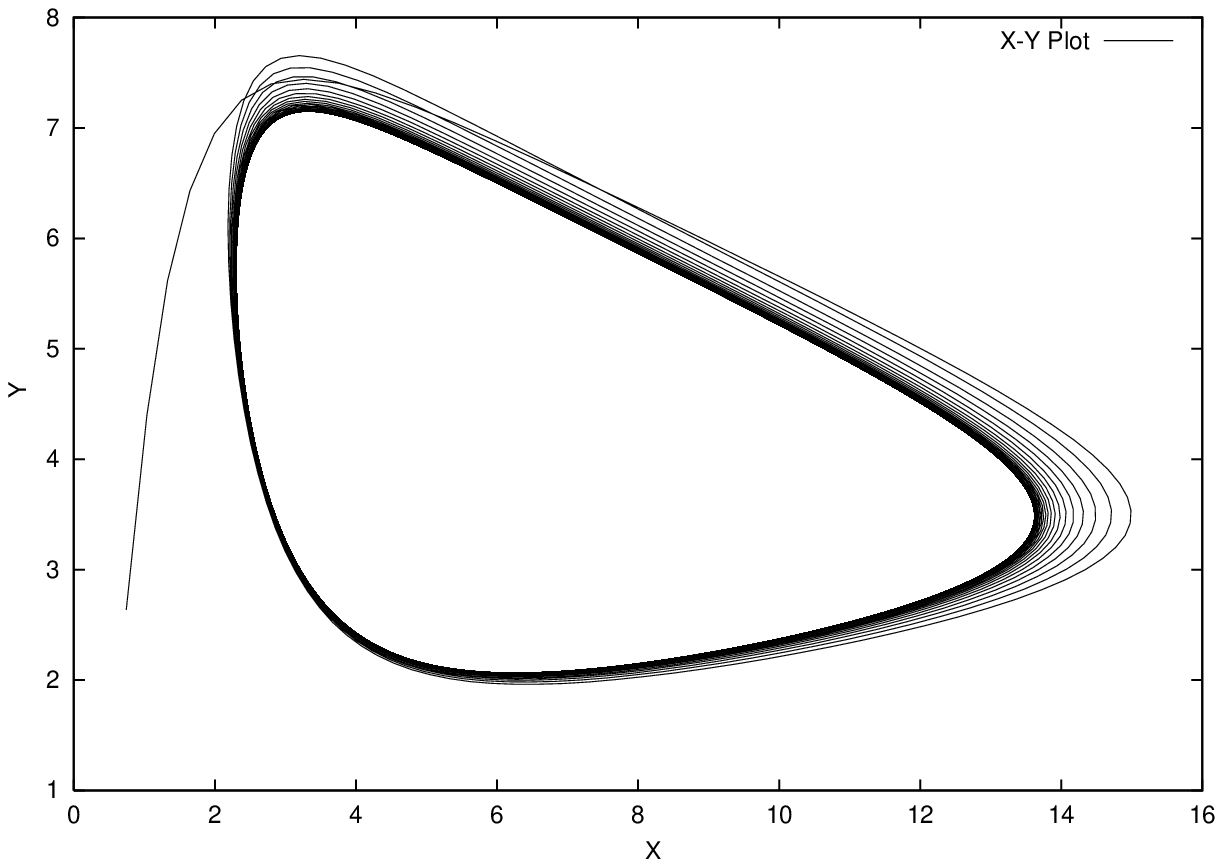}\includegraphics[width=2in]{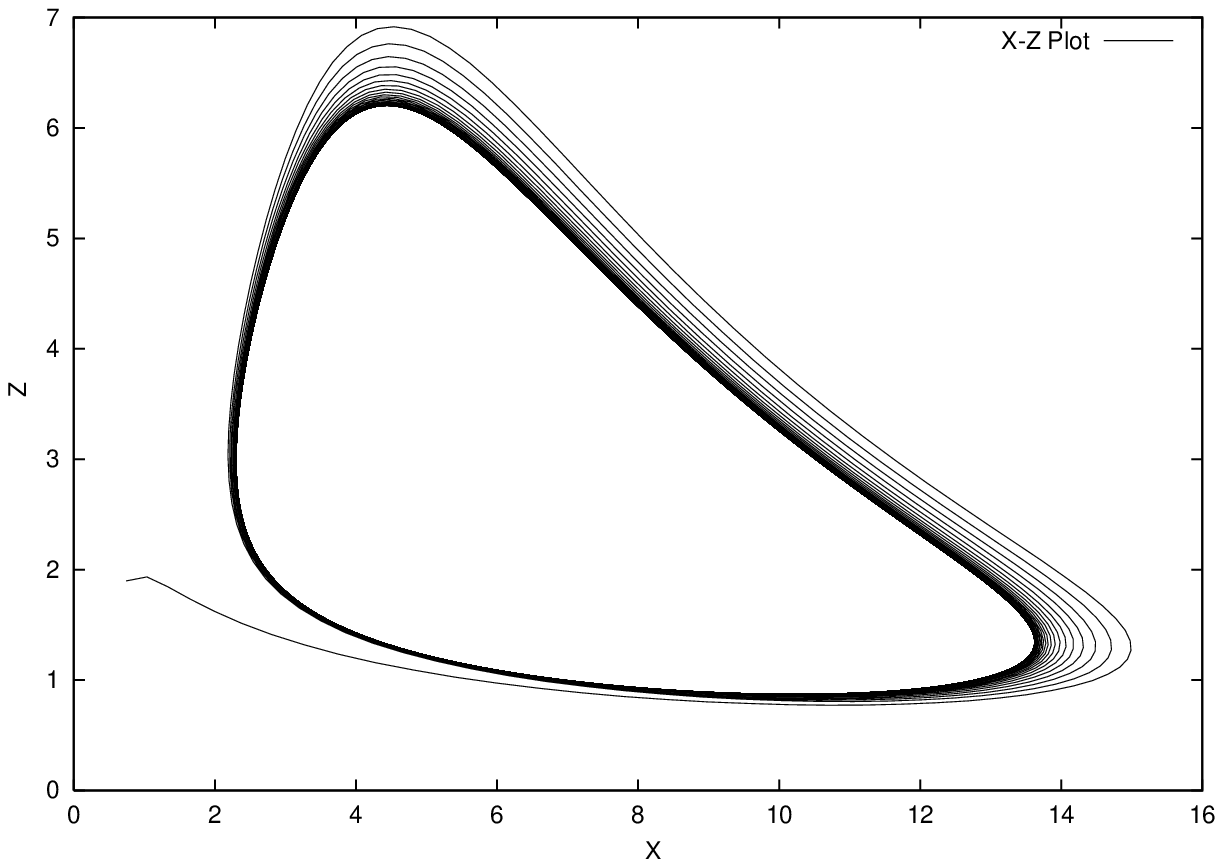}\includegraphics[width=2in]{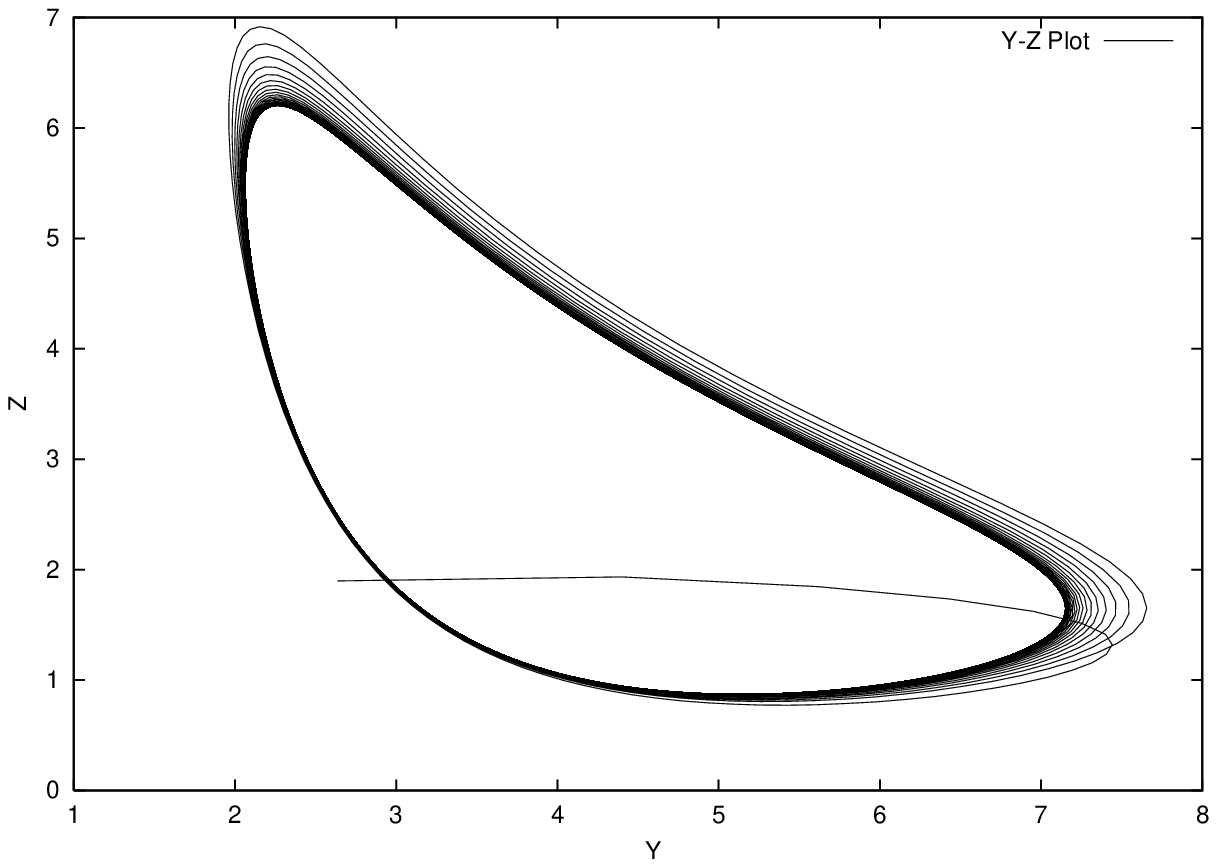}
\par\end{centering}

\begin{centering}
Fig. 3. Different Phase plots corresponding to the oscillatory behaviour
in Fig. 2
\par\end{centering}

\end{figure}

\begin{figure}[H]
\begin{centering}
\includegraphics[width=2.5in]{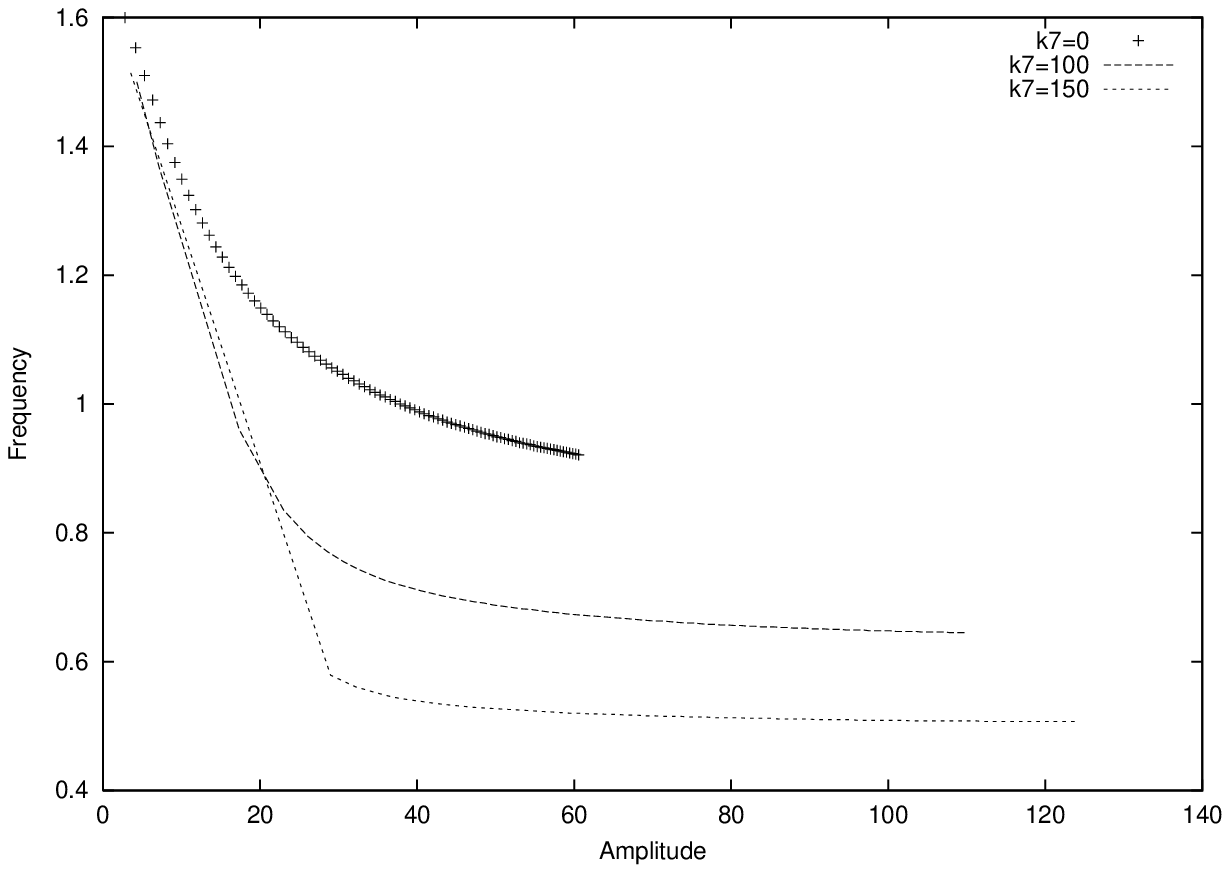}
\par\end{centering}

\centering{}Fig. 4. Amplitude versus Frequency plot when varied $k_{1}$
with different values of positive feedback strength shown in the graph.
The other rate constants are fixed at $k_{2}$=5.730, $k{}_{3}$=331.660,
$k{}_{4}$=3.681, $k{}_{5}$=494.232, $k{}_{7}$=9.168, $K{}_{1}$=1.280,
$K{}_{2}$=0.982, $K{}_{3}$=0.959, $K{}_{4}$=18.882, $n{}_{1}$=2.658,
$n{}_{2}$=2.048, $n{}_{3}$=2.512, n$_{4}$=3.940.
\end{figure}

\begin{figure}[H]
\begin{centering}
\includegraphics[width=2.5in]{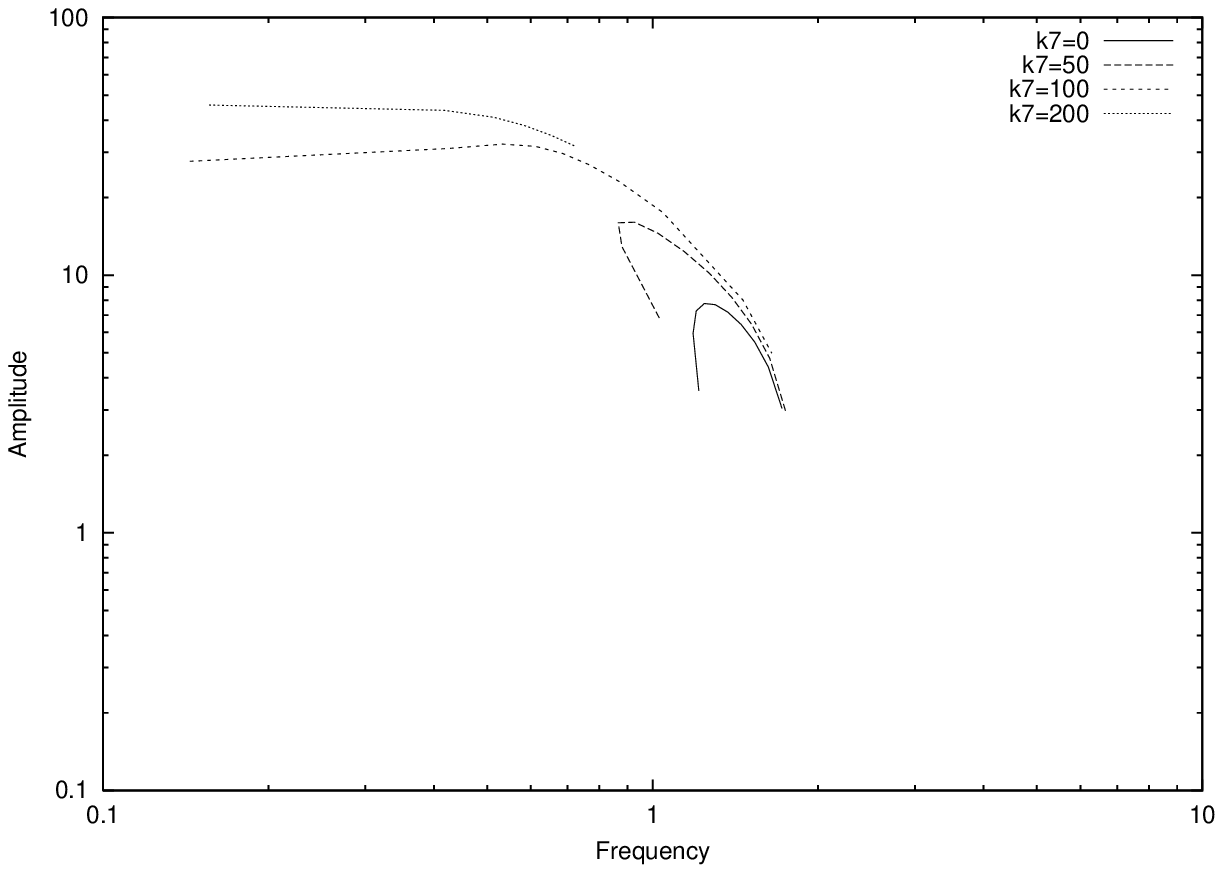}
\par\end{centering}

\begin{centering}
Fig. 5. Frequency versus Amplitude plot with $k{}_{2}$ variation
for different values of positive feedback strength shown in the graph.
The other rate constants are fixed at $k{}_{1}$=266.152, $k{}_{3}$=331.660,
$k{}_{4}$=3.681, $k{}_{5}$=494.232, $k{}_{7}$=9.168, $K{}_{1}$=1.280,
$K{}_{2}$=0.982, $K{}_{3}$=0.959, $K_{4}$=18.882, $n{}_{1}$=2.658,
$n{}_{2}$=2.048, $n{}_{3}$=2.512, n$_{4}$=3.940.
\par\end{centering}

\end{figure}

\begin{figure}[H]
\begin{centering}
\includegraphics[width=2.5in]{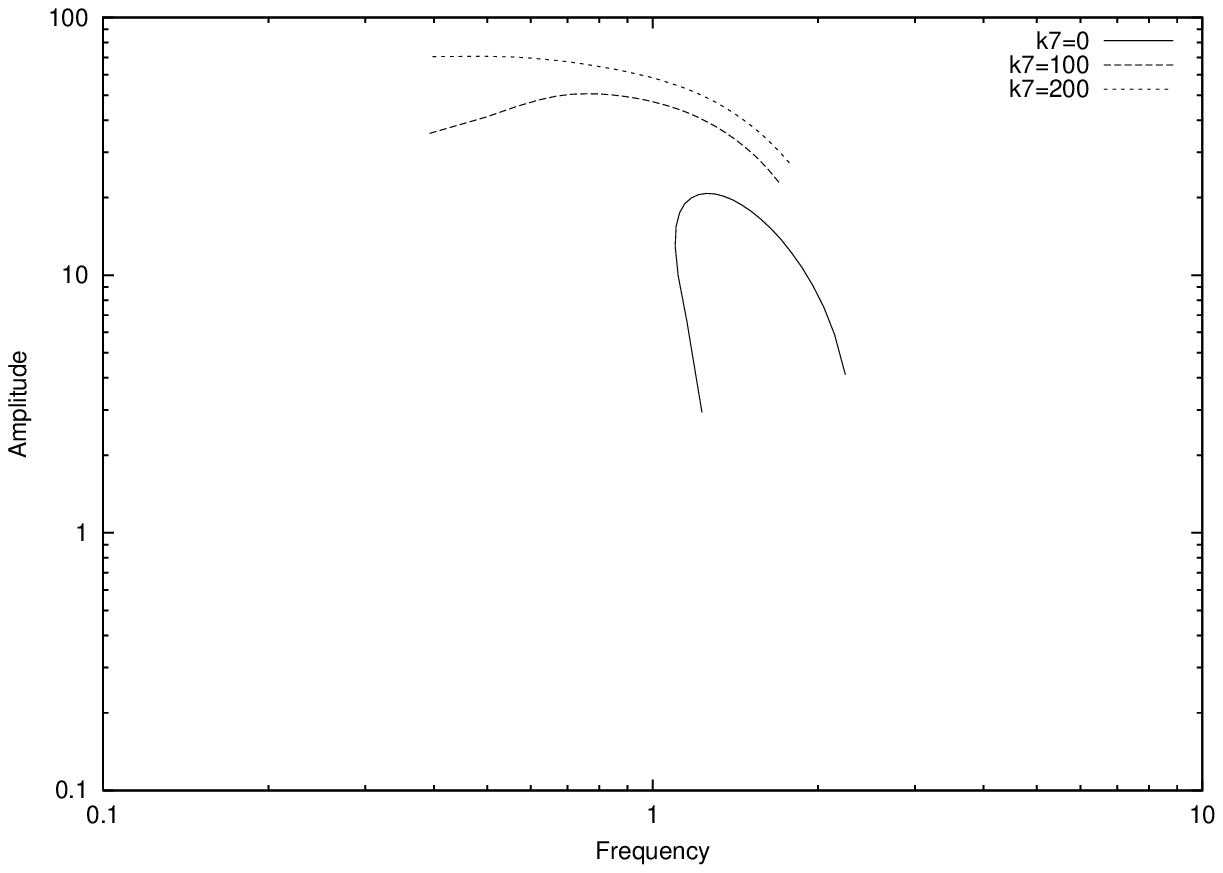}
\par\end{centering}

\centering{}Fig. 6. Frequency versus Amplitude plot with $k{}_{2}$
and $k{}_{1}$(both) variation for different values of positive feedback
strength shown in the graph. The other rate constants are fixed at
$k{}_{3}$=331.660, $k{}_{4}$=3.681, $k{}_{5}$=494.232, $k{}_{7}$=9.168,
$K{}_{1}$=1.280, $K{}_{2}$=0.982, $K{}_{3}$=0.959, $K{}_{4}$=18.882,
$n{}_{1}$=2.658, $n{}_{2}$=2.048, $n{}_{3}$=2.512, $n{}_{4}$=3.940.
\end{figure}

\begin{figure}[H]
\begin{centering}
\includegraphics[width=2.4in]{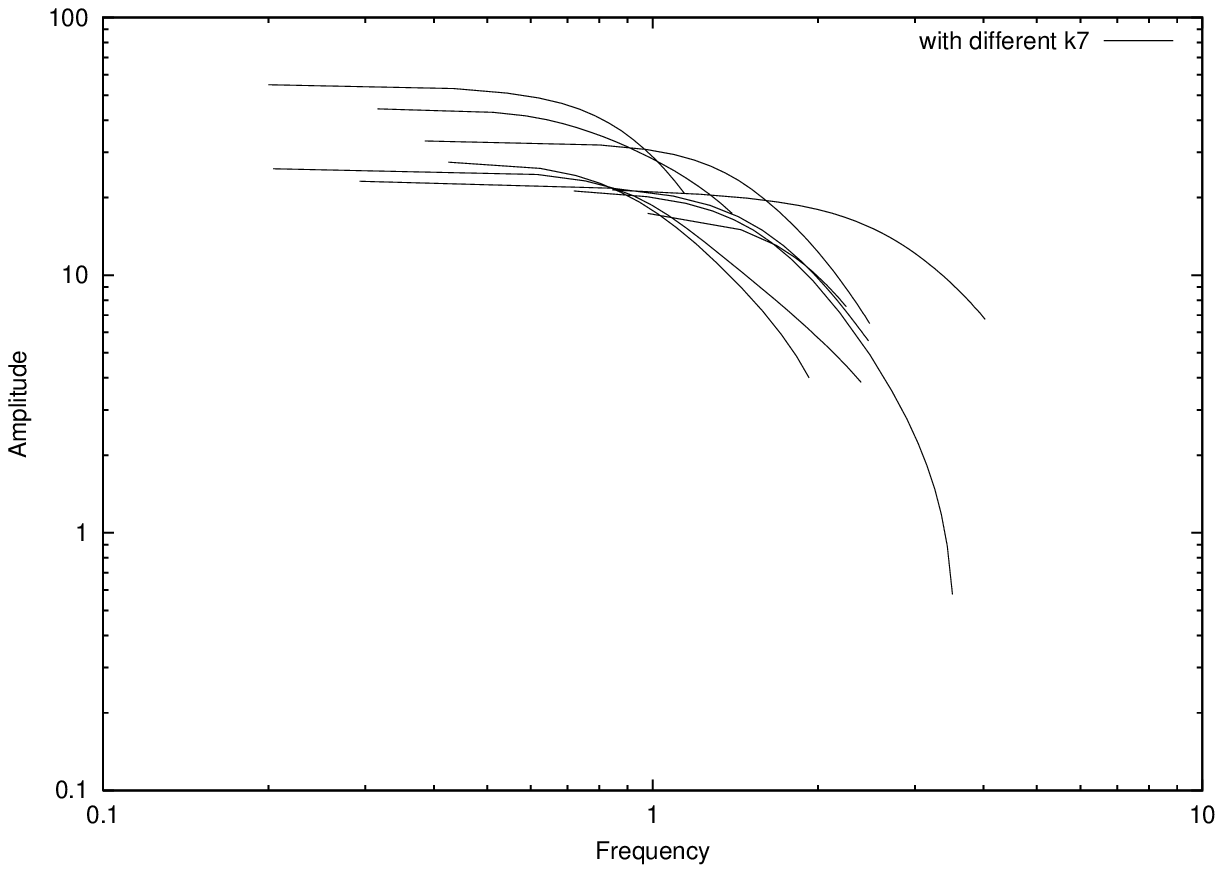}
\par\end{centering}

\centering{}Fig. 7. Frequency versus Amplitude plot with $k{}_{7}$variation.
The other rate constants are fixed at $k{}_{1}$=266.152, $k{}_{2}$=5.730,
$k{}_{3}$=331.660, $k{}_{4}$=3.681, $k{}_{5}$=494.232, $k{}_{7}$=9.168,
$K{}_{1}$=1.280, $K{}_{2}$=0.982, $K{}_{3}$=0.959, $K{}_{4}$=18.882,
$n{}_{1}$=2.658, $n{}_{2}$=2.048, $n{}_{3}$=2.512, $n{}_{4}$=3.940.
\end{figure}

\begin{table}[H]
\caption{Different rate constants with their ranges used to solve the coupled
equation numerically. }

\centering{}%
\begin{tabular}{|c|c|c|}
\hline 
Rate Constant & Value/Range & Description\tabularnewline
\hline 
\hline 
$k_{2}$ & 0-20 & Degradation rate constant\tabularnewline
\hline 
$k_{4}$ & 0-20 & Degradation rate constant\tabularnewline
\hline 
$k_{6}$ & 0-20 & Degradation rate constant\tabularnewline
\hline 
$k_{1}$ & 0-500 & Negative feedback/repression Strength\tabularnewline
\hline 
$k_{3}$ & 0-500 & Negative feedback/repression Strength\tabularnewline
\hline 
$k_{5}$ & 0-500 & Negative feedback/repression Strength\tabularnewline
\hline 
$k_{7}$ & 0-100 & Autocatalytic Positive feedback strength\tabularnewline
\hline 
$n_{1}$ & 2-4 & Hill coefficient of repression\tabularnewline
\hline 
$n_{2}$ & 2-4 & Hill coefficient of repression\tabularnewline
\hline 
$n_{3}$ & 2-4 & Hill coefficient of repression\tabularnewline
\hline 
$n_{4}$ & 2-4 & Hill coefficient of auto-activation\tabularnewline
\hline 
$K_{1}$ & 0-2 & Half maximum value of repressive Hill function at \emph{z}=$K{}_{1}$\tabularnewline
\hline 
$K_{2}$ & 0-2 & Half maximum value of repressive Hill function at x=$K{}_{2}$\tabularnewline
\hline 
$K_{3}$ & 0-2 & Half maximum value of repressive Hill function at y=$K{}_{3}$\tabularnewline
\hline 
$K_{1}$ & 0-20 & Half maximum value of repressive Hill function at x=$K{}_{4}$\tabularnewline
\hline 
\end{tabular}
\end{table}

\section{Conclusion}

We study a gene regulatory network with interlinked positive and negative
feedback loop. The loop produces two different modes of oscillation.
In one mode (mode 1) frequency remains constant over a wide range
amplitude and in other mode (mode 2) the amplitude of oscillation
remains constant over a wide range of frequency. For circadian rhythm,
mode 1 oscillation is very important because organisms try to maintain
a constant frequency of their daily clocks in spite of the variation
of the environmental condition. Mode 2 oscillation is important for
heart beat or cell cycle for which fixed amplitude of oscillations
is very much crucial in different frequency region. Our study reproduces
both features of oscillations in a single gene regulatory network
and show that the negative plus positive feedback loops in gene regulatory
network offer additional advantage. We identified the key parameters/variables
responsible for different modes of oscillation. The network is flexible
in switching between different modes by choosing appropriately the
required parameters/variables. Therefore, gene regulatory networks
with interlinked positive and negative feedback loops work as more
superior oscillator rather than the signaling networks.

\medskip{}

\begin{center}
ACKNOWLWDGWMENT
\par\end{center}

\begin{center}
This work was supported by UGC (ERO) MRP Grant (No. F.PSW-197/11-12(ERO)).
\par\end{center}

\end{document}